\documentstyle[12pt]{article}
\normalbaselines
\begin{document}
\def\square{\hbox{\vrule\vbox{\hrule\phantom{o}\hrule}\vrule}}
\def\wh{\widehat}
\def\ov{\overline}
\renewcommand{\thefootnote}{\fnsymbol{footnote}}
\rightline{
hep-th/9611089
}

\rightline{CAT/95-06}

\hrule
\vskip 2cm
\begin{center}
{\Large{\bf On the spontaneous break down of\\ 
massive gravities in 2+1 dimensions}}\\ [7mm]

{\bf C. Aragone$^{a,}$\footnote{\bf In Memoriam},  
P.J.Arias$^{b,c,}$\footnote{email: parias@tierra.ciens.ucv.ve}
and A.Khoudeir$^{c,}$\footnote{adel@ciens.ula.ve}}\\
 
${}^a${\it Departamento de F\'{\i}sica,
Universidad Sim\'on Bol\'{\i}var, Caracas, Venezuela.}\\[4mm] 
${}^b${\it Grupo de Campos y Part\'{\i}culas, Departamento de F\'{\i}sica, 
Facultad de Ciencias,\\ 
Universidad Central de Venezuela, AP 47270, Caracas 1041-A, Venezuela.}\\[4mm]
${}^c${\it Centro de Astrof\'{\i}sica Te\'orica, Facultad de Ciencias,
 Universidad de los Andes,\\
 AP 26, La Hechicera, M\'erida 5101, Venezuela.}\\ [27mm]

{\bf Abstract}
\end{center}

We show that locally Lorentz invariant, third order, topological massive 
gravity can not be broken down neither to the local diffeomorphism subgroup 
nor to the rigid Poincar\'e group. On the other hand, the recently formulated, 
locally diffeomorphism invariant, second order massive triadic 
(translational) Chern-Simons gravity breaks down on rigid Minkowski space 
to a double massive spin-two system. This flat double massive action is the 
uniform spin-two generalization of the Maxwell-Chern-Simons-Proca system which 
one is left with after U(1) abelian gauge invariance breaks down in the 
presence of a sextic Higgs potential.

\newpage

Topological massive gravity [1] remains a puzzling theory. In spite of being 
third order it is yet unclear whether it is renormalizable [2]. In addition, 
although several types of exact solutions have been found [3], none of them 
are proper Chern-Simons(CS) black holes (or black strings [4]). The black holes 
recently found by Ba\~nados, Teitelboim and Zanelli [5] are three-dimensional 
pure Einstein black holes. They are not specific solutions of topological 
massive gravity.

Another aspect which up to now did not receive much attention concerns about the
possible symmetry breaking process the theory can undergo. This letter aims to
investigate this problem. We study the possibility of breaking down its local
Lorentz invariance and analyze the physical relevance of the two reasonable
models that in such case one is left with.

Since for both types of process the answer will be negative, we then
explore whether topological triadic $CS$-gravity [6] might be spontaneously
broken (to Minkowski space). In this case the answer is positive.

The first process we study arises when one assumes that local Lorentz
invariance is lost by the addition of the diffeomorphism invariant triadic $CS$
term. This action has, then, three constituents: the original third order
Lorentz-$CS$ term $\sim \omega\varepsilon\partial\omega$, minus the second
order Einstein Kinetic term $\sim \omega_{pa}\omega^{ap}$, plus the first order
triadic $CS$ term $\sim e\varepsilon\partial e$ which breaks the local Lorentz
symmetry, since it only possess diffeomorphism invariance. In principle, this
action has a good aspect since it is possible to show it is a pure spin-2 action
which contains two massive excitations of opposite helicities or  two massive
excitations with the same helicity (according to the relative sign  of the
typical massess of the two different CS terms).
We shall see that, however, the energy of this system is not definite 
positive. Consequently this system has not physical relevance. 

Since this process is not allowed we then investigate whether if one takes a 
more simple minded point of view and fully breaks local Lorentz invariance by 
the addition of an algebraic Fierz-Pauli massive term and goes down to flat 
space one obtains a meaningful pure spin-2 action.

We shall see that also in this case the broken system has a pure spin-2 
content which might have either one or three physical poles for its cubic 
propagator, according to the relative value of the two masses involved in the 
original action. Disregarding the case of one physical excitation (it 
implies two additional complex poles) we then analyse the case of the likely 
existence of the three different positive masses.
Also in this case the system shows unbounded energy. Consequently it has no
physical  significance either.

We conclude that Lorentz-CS topological massive gravity cannot be 
spontaneously broken. (We do not foresee any reason why a mixture of the 
triadic-CS term with the Fierz-Pauli one would provide a positive answer).

In view of these negative results for the posibility of breaking down 
standard topological gravity we examine massive vector (triadic) CS-gravity 
[6]. This is a curved theory propagating one massive spin-2 excitation.

We show that in this case, the addition of the Fierz-Pauli (metric) mass term 
to the linearized action (composed by the sum of the three dimensional 
Einstein action $\sim \omega \varepsilon \partial e -\omega^2$ and the triadic CS 
term $\sim e\varepsilon \partial e$) gives rise to a physically relevant pure spin-2 
theory on flat Minkowski space which propagates two spin-2 massive 
excitations of opposite helicities and different masses. Triadic CS 
topological gravity can be broken down to Miskowski space while Lorentz CS 
theory does not allow this type of process.

Now we come to the definitions in order to be able to present the 
technical details.

Three dimensional Einstein action enters in all the theories we consider.Its 
first order form is
$$
E \cdot\equiv \kappa^{-2}<\omega_p{}^a\varepsilon^{prs}\partial_re_{sa}-
2^{-1}e_p{}^a\varepsilon^{prs}\varepsilon_{abc}\omega_r{}^b\omega_s{}^c>
\eqno (1)
$$
where $\kappa$, in units of $[m]^{-1/2}$, is the three dimensional gravitational 
constant. $e_p{}^a$ and $\omega_p{}^a$ constitute the basic triadic 
fields and connections. Middle of the alphabet letters $p,r,\cdots$ name 
world indices while $a,b,\cdots$ denote local Lorentz ones, 
$\varepsilon^{012}=+1$ 
and the flat metric $\eta_{ab}$ has positive signature ($\eta_{00}=-1$). As all
the remaining elementary actions we will introduce in this paper, it is
locally trivial, i.e. the associated field equations tell us that this action
does not propagate local physical excitations.

A simple and useful property of this action is how it looks like when one 
works in a second order formulation. After introducing in eq.(1) 
$\omega_p{}^a$ in 
terms of the triadic fields $e$ as given by the $\omega$-field equations
$$
\varepsilon^{pqr}(\partial_qe_{ra}-\omega_q{}^c\varepsilon_{cba}e_r{}^b)=0, 
\eqno (2)
$$
one is lead to the second order form of $E$. It turns out to be 
$$
E=2^{-1}\kappa^{-2}<\varepsilon^{prs}\varepsilon_{abc}e_p{}^a\omega_r{}^b(e)
\omega_s^c(e)>,\eqno (3)
$$
which in the linearized case, takes the typical Fierz-Pauli form 
$\sim 2^{-1}\omega_{pa}\omega_{ap}-2^{-1}\omega^2$ in terms of the (now non 
independent) variables $\omega$. Topological massive gravity (which from now on
we call Lorentz-CS gravity)  needs the presence of another locally trivial
action: the third order,  locally conformal and Lorentz invariant Chern-Simons
term
$$
L\cdot\equiv (2\mu_1\kappa^{-2})^{-1}<\omega_p{}^a\varepsilon^{prs}
\partial_r
\omega_{sa}-2(3)^{-1}\varepsilon^{prs}\varepsilon_{abc}\omega_p{}^a
\omega_p{}^b\omega_s{}^c>
\eqno (4)
$$
where $\omega = \omega (e)$ as given by eq.(2). Its action is $L-E$.

Triadic CS-gravity [6] on the other hand is defined by adding to E the first 
order, diffeomorphism invariant triadic-CS term:
$$
T\cdot\equiv 2^{-1}\mu_2\kappa^{-2}<e_p{}^a\varepsilon^{prs}\partial_re_{sa}>.
\eqno (5)
$$
The associated full second order action is $E+T$.

We want to investigate whether Lorentz-CS gravity can be broken by a term 
$\sim$ T; i.e. we wonder whether $L-E+T$ is a satisfactory spin-2 theory 
having 
a mass spectra corresponding to some spliting of the initial mass $\mu$ in 
two different (but closely related) masses.

A reasonable insight on the physical significance of this proposal can be 
obtained by analysing the behaviour of the associated third-order linearized 
theory on flat Minkowski space. We start considering its first order action $S_0$. 
It is the quadratic part of $L-E+T$ when we expand 
$e_p{}^a=\delta_p^a+\kappa h_p^a$ in terms of $\kappa$
$$
S_0\cdot =(2\mu )^{-1}<\omega_p{}^a\varepsilon^{prs}\partial_r\omega_{sa}>-
2^{-1}<\omega_{pa}\omega^{ap}-\omega^2>-2^{-1}m<h_p{}^a\varepsilon^{prs}
\partial_r h_{sa}>
$$
$$
+<\lambda_p{}^a\varepsilon^{prs}(\partial_rh_{sa}-\omega_r{}^b
\varepsilon_{bsa})>.\eqno (6)
$$
Its equivalent third order version arises from introducing the values of 
$\omega =\omega (h)$ (obtained from variations of the $\lambda$'s) into 
$S_0$.

Independent variations of $\omega$, $h$, $\lambda$ yield the triplet of field 
equations ($FE$).
$$
E^p{}_a\cdot =\mu^{-1}\varepsilon^{prs}\partial_r\omega_{sa}-\omega_a{}^p+
\delta_a{}^p\omega-\lambda_a{}^p+\delta_a{}^p\lambda =0,\eqno (7)
$$
$$
F^p{}_a=-m\varepsilon^{prs}\partial_rh_{sa}+\varepsilon^{prs}\partial_r
\lambda_s{}^a=0,\eqno (8)
$$
and
$$
G^p{}_a\cdot =\varepsilon^{prs}\partial_rh_{sa}-\omega_a{}^p+\delta_a{}^p
\omega =0.\eqno (9)
$$

Considering the lower spin sector of these eqs., i.e. computing 
$E\cdot \equiv E^p{}_p$, $F$, $G$, $\partial_pE^p{}_a,\cdots$ and 
$\varepsilon_{pab}E^{pa}\equiv \cdot \check{E}_b$, $\check{F}_b$, $\check{C}_b$ 
it is straightforward to see that this system only propagates spin-2 
excitations.

Both, the spin-1 
$\varepsilon_{pab}\omega^{pa},\cdots,\varepsilon_{pab}\lambda^{pa},
\partial_p\omega_{pa},\partial_p\lambda_{pa}$ 
and the scalar sector of $\omega ,h, \lambda$ vanish in the harmonic gauge 
$\partial_ph_{pa}=0$.

Projection of the $FE$ (7) (8) (9) upon the spin-2$^+$ (spin-2$^-$) subspaces 
using the pseudo-seudospin-2$^\pm$ projectors [7], gives
$$
(X-1)\omega^{T+}-\lambda^{T+}=0\  \ ,\  \ -mXh^{T+}+X\lambda^{T+}=0\  \ , \  \ 
\omega^{T+}=Xh^{T+} \eqno (10)
$$
where $X=\mu^{-1}\square^{1/2},\mu =1$. $m$ means the dimensionless relation 
$m\mu^{-1}$ and $h^{T+}$ denotes the spin-2$^+$ part of $h_{pa}$.

The inverse propagator is therefore
$$
\Delta^+(X)=X[X(X-1)-m].\eqno (11)
$$

There is a positive mass $m=2^{-1}+(4^{-1}+m)^{1/2}$ in the spin-2$^+$ sector. 
Similarly, since $\Delta^-(X)=X[X(X+1)-m]$ we might have a spin-2$^-$ 
excitation with mass $m^-=-2^{-1}+(4^{-1}+m)^{1/2}$.

We want to see whether this system has its energy bounded from below (or not). 
It will be shown that, independently of the sign of $m$, the light-front (LF) 
generator is unbounded and consequently action (6) is physically meaningless, 
in spite of the fact that, from a covariant point of view, the system (7), 
(8), (9), seems to propagate two spin-2 decoupled excitations.

In order to have this, we calculate the value of the LF-generator of action 
(6) in terms of its two unconstrained variables $\omega_{vv}$ and 
$\lambda_{vv}$. Light front coordinates $(u,v)$ are defined by
$$
\eta^{11}=1=-\eta^{uv},\  \ u\cdot =2^{-1/2}(x^0-x^2),\  \ 
v\cdot =2^{-1/2}(x^0+x^2),\  \ \varepsilon^{1vu}=+1.\eqno (12)
$$
Time derivatives are written $\partial_u f=\dot{f}$ and the LF-spacelike ones 
are denoted $\partial_vf=f'$

One starts from the covariant expressions (6) of $S_0$ and express this action 
in terms  of the 27 $LF$-field components $\omega_{uu}\equiv\cdot \omega_u$, 
$\omega_{uv}$, $\omega_{vu}$, $\omega_v\cdot \equiv\omega_{vv}$, 
$\omega_1\cdot \equiv\omega_{11}$, $\omega_{1u}$, $\omega_{u1}$, $\omega_{1v}$ and 
$\omega_{v1},\cdots ,\lambda_u,\lambda_{uv},\cdots ,\lambda_{1v},\lambda_{v1}$.

It is inmediate to realize that $\omega_{ua},h_{ub},\lambda_{uc}$ are 
multipliers associatted with nine differential constraint equations which 
can be solved, providing the values of $\omega_{1a},h_{1b},\lambda_{1c}$ as 
functions of the remaining nine intermediate variables 
$\omega_{va},h_{vb},\lambda_{vc}$. Their solution is:
$$
\wh{\omega}_{1v}=(\partial_1+1)\wh{\omega}_v+\wh{\lambda}_v
\  \ ,\  \ \wh{h}_{1v}=\partial_1\wh{h}_v+\wh{\omega}_v,\eqno (13a,b)
$$
$$
\wh{\lambda}_{1v}=\partial_1\wh{\lambda}_v+m\wh{\omega}_v,
\eqno (13c)
$$
$$
\omega_1=\partial_1\wh{\omega}_{v1}+\wh{\omega}_{1v}+
\wh{\lambda}_{1v}\  \ ,\  \ h_1=\partial_1\wh{h}_{v1}+
\wh{\omega}_{1v},\eqno (14a,b)
$$
$$
\lambda_1=\partial_1\wh{\lambda}_{v1}+m\wh{\omega}_{1v},\eqno (14c)
$$
$$
\omega_{1u}{}'=(\partial_1-1)\omega_{vu}-\lambda_{vu}+\partial_1
\wh{\omega}_{v1}+\partial_1\wh{\lambda}_{v1}+(m+1)
\wh{\omega}_{1v}+\wh{\lambda}_{1v}\eqno (15a)
$$
$$
h_{1u}{}'=\partial_1h_{vu}-\omega_{vu}+\partial_1\wh{\omega}_{v1}+
\wh{\omega}_{1v}+\wh{\lambda}_{1v}\eqno (15b)
$$
$$
\lambda_{1u}{}'=\partial_1\lambda_{vu}-m\omega_{vu}+
m\partial_1\wh{\omega}_{v1}+m\wh{\omega}_{1v}+
m\wh{\lambda}_{1v}\eqno (15c)
$$
where we introduced redefinitions like $\omega_{v1}\equiv\cdot
\wh{\omega}'_{v1}$,  $\omega_{v}\equiv\cdot \wh{\omega}''_{v}$, 
$\omega_{1v}\equiv\cdot \wh{\omega}'_{1v}$ for the 
three sets of variables $\omega$, $\lambda$, $h$.

In principle, the intermediate expression of $S_0$ obtained in terms of the 
nine intermediate variables $\omega_{va},h_{vb},\lambda_{vc}$ might have 
$\omega_{vu},h_{vu},\lambda_{vu}$ in the dynamical germ (the piece of 
$S_0\sim p\dot{q}$).

However it turns out after using eqs. (13), (14) that these three variables 
are not present in this part of the action. While $\omega_{vu}$ and 
$\lambda_{vu}$ constitute two additional Lagrange multipliers, $h_{vu}$ has 
totally disappeared.

Independent variations of $\omega_{vu},\lambda_{vu}$ lead to the final two 
differential constraints of $S_0$. Their solution shows the symmetry of the
$1v$-components $\omega_{1v},\lambda_{1v}$, i.e. 
$$
\wh{\omega}_{v1}=\wh{\omega}_{1v}\  \ ,\  \ 
\wh{\lambda}_{v1}=\wh{\lambda}_{1v}\eqno (16a,b)
$$

Now it is immediate to obtain the unconstrained form of the evolution 
generator $G$ of action $S_0\sim p\dot{q}-G$. Since, at the initial stage 
when one writes down $S_0$ in terms of the LF-variables, $G$ had the form:
$$
G=<(\wh{\omega}_{v1}+\wh{\lambda}_{v1})\omega_{1v}{}'+\wh{\omega}_{v1}
\lambda_{1u}{}'>;\eqno (17)
$$
it is straightforward to realize that, after insertion of the values (15) of 
$\omega_{1u}{}',\lambda_{1u}{}'$ in it, $G$ becomes:
$$
G=<[(m+1)\wh{\omega}_{1v}+\wh{\lambda}_{1v}]^2-m^2\wh{\omega}^2_{1v}>.
\eqno (18)
$$
This explicitly shows that the generator is a non semidefinite positive 
quadratic expression. Consequently the unconstrained reduced form of 
$S_0$, even written in terms of the unique two gauge-invariant variables 
$\wh{\lambda}_v,\wh{\omega}_v$, does not have physical relevance. 
One can say that the presence of both types of $CS$ terms is inconsistent. This 
situation is peculiar of Lorentz-$CS$ gravity (there is no analogous third
order  $CS$ theory for vector fields).

Since $LCS$-gravity can not be broken through a triadic $CS$ type of term we
now consider the possibility of a harder type of breaking induced by the
presence, in flat Minkowsky space, of a Fierz-Pauli mass term. In order to
investigate this  possibility we examine the linearized system. It consists of
$$
S_3\cdot\equiv L^Q-E^Q+2^{-1}m^2\varepsilon <h_{pa}h^{ap}-h^2>\eqno (19)
$$
where $L^Q$, and $E^Q$ are the quadratic parts (in terms of $\kappa$) of the 
exact curved actions (4) and (1) respectively, 
$e_{pa}\equiv \eta_{pa}+\kappa h_{pa}$.

It is convenient to start from a first order system equivalent to (19). It 
reads ($\omega \to \kappa\omega$)

\vspace{5mm}
$$
S_1\cdot =(2\mu )^{-1}<\omega_p{}^a\varepsilon^{prs}\partial_r\omega_{sa}>
-2^{-1}<\omega_p{}^a\omega_a{}^p-\omega^2>
$$
$$
+2^{-1}m^2\varepsilon <h_{pa}h^{ap}-h^2>+
<\lambda_p\varepsilon^{prs}(\partial_rh_s{}^a-\omega_r{}^b\varepsilon_{bsa})>
\eqno (20)
$$
where $\lambda ,\omega ,h$ are three sets of independent variables and 
$\varepsilon =\pm 1$.

Calculations are still simpler using new dimensionless variables 
$x^r_{new}=mx^r$, $h^{pa}=m^{1/2}h^{pa}_{new}$, 
$\omega^{pa}=m^{3/2}\omega_{new}^{pa}$, 
$\lambda^{pa}=m^{3/2}\lambda_{new}^{pa}$. (Without risk of ambiguity the 
subscript new is pressumed in all variables from now on).

Independent variations with respect to them yield three sets of field
equations$(FE)$ 
$$
\delta S_1/\delta\omega \to E^p{}_a=0\  \ ,\  \ 
\delta S_1/\delta h\to F^p{}_a=0\  \ ,\  \ 
\delta S_1/\delta\lambda \to G^p{}_a=0. \eqno (20a,b,c)
$$

Straightforward calculations with 
$E^p{}_p,\partial_pE^p{}_a,\partial^a\partial_pE^p{}_a,
\partial^a\partial_pE^p{}_a,\varepsilon^{pba}E_{pa}$ show that all lower spin 
variables contained in $\omega_p{}^a,h_p{}^a,\lambda_p{}^a$ vanish on the FE. 
(There is no need of decomposing triadic variables into their irreducible 
symmetric and antisymmetric parts). Then, the system generated by action (20) is
a pure spin-2 system.

In the presence of a consistent current $j_{pa}$  
($\varepsilon^{pab}j_{pa}=0$, $\partial_pj_{pa}=0$) the triplet of $FE$ reads
$$
E^{T\ov{pa}}=\frac{1}{2\mu}
(\varepsilon_p{}^{rs}_\cdot \partial_r\omega^T_{\ov{sa}}+\varepsilon_a{}^{rs}
\partial_r\omega_{\ov{sp}}^T)-\omega_{\ov{pa}}^T-\lambda_{\ov{pa}}^T=0,
\eqno (21)
$$
$$
F^{T\ov{pa}}=\frac{1}{2}
(\varepsilon_p{}^{rs}_\cdot \partial_r\lambda^T_{\ov{sa}}+\varepsilon_a{}^{rs}
\partial_r\lambda_{\ov{sp}}^T)-\varepsilon h_{\ov{pa}}^T=-j_{pa}^T,
\eqno (22)
$$
$$
G^{T\ov{pa}}=\frac{1}{2}
(\varepsilon_p{}^{rs}_\cdot \partial_r h^T_{\ov{sa}}+\varepsilon_a{}^{rs}
\partial_r h_{\ov{sp}}^T)-\omega_{\ov{pa}}^T=0.
\eqno (23)
$$
in terms of the symmetric traceless transverse parts 
($h^T_{\ov{pp}}=0=\partial_ph^T_{\ov{pa}}$) of the initial non-symmetric
variables $\omega_p{}^a,h_p{}^a,\lambda_p{}^a$.

The last step in this quick preliminary covariant analysis consists in using 
the simple helicity projectors $P_2^\pm$ [7] for separating, given a symmetric 
traceless, transverse second order rank tensor, its two spin-2$^+$ and 
spin-2$^-$ components.

For second rank symmetric traceless transverse tensors $h^T_{\ov{pa}}$ we 
have
$$
h^{T+}_{\ov{pa}}-h^{T-}_{\ov{pa}}\cdot =\{ (P^+_2-P^-_2)h^T\}_{\ov{pa}}=
2^{-1}\square^{-\frac{1}{2}}(\varepsilon_p{}^{rs}_\cdot 
\partial_rh^T_{\ov{sa}}+\varepsilon_a{}^{rs}\partial_rh^T_{\ov{sp}}), 
\eqno (24)
$$
$$
h^{T+}_{\ov{pa}}+h^{T-}_{\ov{pa}}=\{ (P^+_2+P^-_2)h^T\}_{\ov{pa}}=
h^T_{\ov{pa}}.\eqno (25)
$$

Projection of eqs. (21)$\cdots$(23) (on the spin-2$^+$ sector) leads to 
($\square^{1/2}=X$)
$$
Xh^{T+}=\omega^{T+}\  \ ,\  \ \mu^{-1}X\omega^{T+}-\omega^{T+}=
\lambda^{T+}\  \ ,\  \ X\lambda^{T+}+\varepsilon h^{T+}=-j^{T+}.
\eqno 26(a,b,c)
$$

It is immediate to calculate the inverse propagator of this system. It is
$$
\Delta^+(X)\cdot =X^2(\frac{X}{\mu}-1)+\varepsilon .\eqno (27)
$$

\noindent $(\Delta^-(X)=\Delta^+(-\mu ,X))$.The mass spectrum of our system are the
positive solutions of  $\Delta^+(X)$ and $\Delta^-(X)$. The analysis of this
cubic equation shows that, for $\mu>0$, the more  interesting case will be when
$\varepsilon =+1$. In this case we might have two positive masses of helicity
$s=2^+$ and one positive mass for the $s=2^-$  excitation. (When $\varepsilon
=-1$ we have complex poles in addition to  only one positive mass).

We explain now why even in what seems to be the more interesting case, 
$\varepsilon =+1$, the system is unphysical, the reason being that its energy 
is unbounded.

The proof consists in obtaining the unconstrained action in terms of the 
three spin-2 independent variables this system has, and then looking at the
explicit form of the energy. We perform the reduction process in the $LF$
coordinates, see eqs(12).The initial system (20) has $3\times 9=27$ independent
variables. Again for  simplicity in the case of two equal indices we skip one
of them, i.e.  $h_{uu}=h_u,h_{vv}=h_v,h_{11}=h_1$. In terms of these $LF$
variables action (20) looks like $$
S_1=\mu^{-1}<\omega_{1u}\dot{\omega}_v+\omega_{1v}\dot{\omega}_{vu}-
\omega_1\dot{\omega}_{v1}>+<h_{1u}\dot{\lambda}_v+h_{1v}\dot{\lambda}_{vu}-
h_1\dot{\lambda}_{v1}>+
$$
$$
+<\lambda_{1u}\dot{h}_v+\lambda_{1v}\dot{h}_{vu}-\lambda_1\dot{h}_{v1}>+
<\omega_{u}{\cal{C}}^{u}+\omega_{uv}{\cal{C}}^v+\omega_{u1}{\cal{C}}^1>+
$$
$$
+<h_{ua}{\cal{D}}^a>+<\lambda_{ub}{\cal{E}}^b+<\omega_{v1}\omega_{1u}-
\omega_1\omega_{vu}>+
$$
$$
+\varepsilon<h_1h_{vu}-h_{v1}h_{1u}>+<\lambda_{v1}\omega_{1u}-\lambda_1
\omega_{vu}+\lambda_{1u}\omega_{v1}-\omega_1\lambda_{vu}>\eqno (28)
$$
where we explicitly see the presence of, at least, 9 differential constraints 
${\cal{C}}^a,{\cal{D}}^a,{\cal{E}}^a=0$ associated with the 9 Lagrange multipliers 
$\omega_{ua},h_{ua},\lambda_{ua}$. They allow to obtain 
$\omega_{1a},h_{1a},\lambda_{1a}$ in terms of the nine variables 
$\omega_{va},h_{va},\lambda_{va}$. The latter set contains more variables
than the ones we would expect for a system that describes three independent
massive degrees of freedom in 2+1 dimensions.

It turns out to be convenient to introduce the new variables: 
$\wh{\lambda}_v,\wh{\omega}_v,\wh{h}_v,
\wh{\lambda}_{v1},\wh{\omega}_{v1},\wh{h}_{v1},
\wh{\lambda}_{1v},\wh{\omega}_{1v},\wh{h}_{1v}$ defined by
$$
\wh{\lambda}_v{}''\cdot \equiv\lambda_v\  \ ,\  \ 
\wh{\lambda}_{v1}{}'\cdot\equiv\lambda_{v1}\  \ ,\  \ 
\wh{\lambda}_{2v}{}'\cdot \equiv\lambda_{1v}\  \ ,\  \ \cdots \eqno (29)
$$
and similar ones exchanging $\lambda$ for $h,\omega$. As usually happens in 
light-front coordinates the 9 constraints are linear ordinary differential
equations  in the ``space-like'' variable $v$. They can be solved in terms
of  $\wh{\lambda}_v,\wh{\lambda}_{v1},\lambda_{vu},
\wh{\omega}_v,\wh{\omega}_{v1},\omega_{vu},\wh{h}_v,\wh{h}_{v1}$ and $h_{vu}$
$$
\wh{\omega}_{1v}=(\partial_1+\mu )\wh{\omega}_v+\mu \wh{\lambda}_v\  \ ,\  \ 
\omega_{1u}{}'=(\partial_1-\mu)\omega_{vu}-\mu\lambda_{vu}+\mu\omega_1+
\mu\lambda_1 \eqno (30a,b)
$$
$$
\omega_1=\partial_1\wh{\omega}_{v1}+\mu \wh{\omega}_{1v}+
\mu \wh{\lambda}_{1v}\  \ ,\eqno (30c)
$$
$$
\wh{\lambda}_{1v}=\partial_1\wh{\lambda}_v -\varepsilon_p\wh{h}_v\  \ ,\  \ 
\lambda_{1v}{}'=\partial_1\lambda_{vu}-\varepsilon_ph_1+
\varepsilon_ph_{vu}\  \ ,\eqno (31a,b)
$$
$$
\lambda_1=\partial_1\wh{\lambda}_{v1}-\varepsilon_p\wh{h}{1v}\  \ ,\eqno (31c)
$$
$$
\wh{h}_{1v}=\partial_1\wh{h}_v -\wh{\omega}_v\  \ ,\  \ 
h_{1u}{}'=\partial_1h_{vu}-\omega_1-\omega_{vu}\  \ ,\eqno (32a,b)
$$
$$
h_1=\partial_1\wh{h}_{v1}-\wh{\omega}{1v}\  \ .\eqno (32c)
$$
Insertion of these values of $\lambda_{1a},\omega_{1b},h_{1c}$ into $S_1$ 
yields, in principle, an apparently unconstrained functional in terms of 
$\lambda_{va},\omega_{vb},h_{vc}$, as expected.

Something interesting happens. $S_1$ has the form $\sim p\dot{q}-G$. The 
dynamical germ (the part of $S_1\sim p\dot{q}$) does not depend upon 
$h_{vu},\omega_{vu},\lambda_{vu}$. These variables only appear linearly in 
the light-front generator $G$ of the system. We have found the remaining three
Lagrange  multipliers of the system. Independent variations with respect to
them  provide the three additional constraints ${\cal{H}}_h,{\cal{H}}_\omega ,
{\cal{H}}_\lambda$ needed in order to have a final $S_1$ fully unconstrained. 
Consequently $S_1$ can be cast as a functional of the three independent physical
variables  $\wh{\lambda}_v,\wh{\omega}_v,\wh{h}_v$,
$$
{\cal{H}}_h\sim \delta S_1/\delta h_{vu}=0\to \wh{\omega}_{v1}=-\varepsilon \mu 
\wh{\lambda}_v-\varepsilon (\partial_1+\mu )\wh{\omega}_v=-\varepsilon 
\wh{\omega}_{1v},\eqno (33)
$$
$$
{\cal{H}}_\omega \sim \delta S_1/\delta \omega_{vu}=0\to \wh{h}_{v1}=
-\varepsilon \wh{\omega}_v-\varepsilon \partial_1 \wh{h}_v=-\varepsilon 
\wh{h}_{1v},\eqno (34)
$$
$$
{\cal{H}}_\lambda \sim \delta S_1/\delta \lambda_{vu}=0\to \wh{\lambda}_{v1}=
\omega_h{\lambda}_{1v} +(1+\varepsilon )\wh{\omega}_{1v}.\eqno (35)
$$

The final step is to introduce these values in $S_1$. In turns out that the 
unconstrained light-front generator has the diagonal, non definite positive form
$$
<G>=<\wh{\omega}_{1v}^2+[\mu (\wh{\omega}_{1v}+\wh{\lambda}_{1v})-\varepsilon 
\wh{h}_{1v}]^2-\varepsilon^2\wh{h}_{1v}^2>.\eqno (36)
$$
The system, having an unbounded light-front-generator, does not have physical
interest.

Thinking in terms of our initial Lorentz $CS$ gravity, we can say that this 
theory can not be broken down by a Fierz-Pauli mass term.

It is worth pointing out what happens if we do not have the Fierz-Pauli-mass
term.  This is linearized Lorentz $CS$ standard topological gravity. Following
along the same lines, one finds that $\wh{\omega}_{1v}=0$,  $\varepsilon \to
0$, and the generator becomes the non negative expression 
$$
<G>^{lim}_{LCS}=\mu^2<\wh{\lambda}_{1v}^2>.\eqno (37)
$$
In this case the constraints tell us that 
$\wh{\lambda}_{1v}=\partial_1\wh{\lambda}_v$ and, futhermore that 
$\wh{\lambda}_v=-(\mu^{-1}\partial_1+1)\wh{\omega}_v$. Choosing 
$h=2^{1/2}\partial_1\wh{\omega}_v$ as the basic dynamical variable the 
unconstrained form of linearized $LCS$ gravity becomes
$$
S_{LCS}^{lim}=<h'\dot{h}>-2^{-1}<h[-\partial_1^2+\mu^2]h>,\eqno (38)
$$
as expected.

Now we apply the same methods to investigate the physical relevance of 
spontaneously breaking triadic CS topological gravity whose exact 
$S_{TCS}=E+T$ action was recently analyzed [6].

The symmetry breaking process we imagine leads to consider, on flat Minkowski 
space, the quadratic system defined by
$$
S_2:=<\omega_p{}^a\varepsilon^{prs}\partial_rh_{sa}>-2^{-1}<\omega_{pa}\omega^{ap}-
\omega^2>+2^{-1}\mu<h_p{}^a\varepsilon^{prs}\partial_rh_{sa}>
$$
$$
-2^{-1}m^2<h_{pa}h^{ap}-h^2>.\eqno (39)
$$
Although this is the first order form of the system, independent variations 
of $\omega$ lead to its quadratic second order expressions in terms of the 
weak field $h_{pa}$.

Independent variations of $h,\omega$ lead to the set of two field equations:
$$
\delta S/\delta h_p{}^a\to E^p{}_a\cdot =\varepsilon^{prs}\partial_r
\omega_{sa}+\mu \varepsilon^{prs}\partial_rh_{sa}-m^2(h_a{}^p-\
\delta_a{}^ph)=0,\eqno (40)
$$
$$
\delta S/\delta \omega_p{}^a\to F^p{}_a\cdot =\varepsilon^{prs}
\partial_rh_{sa}-\omega_a{}^p+\delta_a{}^p\omega =0.\eqno (41)
$$
It is straightforward to see this system has a pure spin-2 content.

Taking traces $E^p{}_p,F^p{}_p$, antisymmetric parts $\varepsilon_{pab}E^{pa}$, 
$\varepsilon_{pab}F^{pa}$ and divergences 
$\partial_pE^p{}_a,\partial_pF^p{}_a$ of the field equations one is lead to 
observe the vanishing of all the scalar and vector parts of 
$h_{pa},\omega_{pa}$ ($h=\omega =0,\varepsilon_{pab}h^{pa}=\varepsilon_{pab}
\omega^{pa}=0$, $\partial_ph_{pa}=0=\partial_p\omega_{pa}$).

There is no need of using the standard symmetric and antisymmetric components 
of $h_{pa},\omega_{pa}$. It is unconvenient. The symmetric parts of equations 
(40),  (41) have the form: $$
X(h^{T+}-h^{T-})=\omega^{T+}+\omega^{T-},\eqno (42)
$$
$$
X(\omega^{T+}-\omega^{T-})+\mu X(h^{T+}-h^{T-})=h^{T+}+h^{T-},\eqno (43)
$$
where we are using dimensionless variables ($\sim m=1$). The characteristic 
equations for each spin-2$^\pm$ component are 
$\Delta_\pm(X)=X^2\pm \mu X-1$. The system has positive masses 
$m^\pm=\mp 2^{-1}\mu +(1+4^{-1}\mu^2)^{1/2}$ for the respective $h^{T+}$ and 
$h^{T-}$ spin-2 excitations.

The viability of this system will emerge from considering the associated 
unconstrained action and showing that its energy is bounded from below.

We shall see this is true. $S_2$ is a physical theory. Indeed it is the 
peculiar $CS$ splitting of the mass degeneracy of double massive, three 
dimensional Einstein-Fierz-Pauli action.

In terms of the light-front variables $S_2$ takes the initial form
$$
S_2=<h_{1u}\dot{\omega}_v+h_{1v}\dot{\omega}_{vu}-h_1\dot{\omega}_{v1}>+
<(\omega_{1u}+\mu h_{1u})\dot{h}_v+(\omega_{1v}+\mu h_{1v})\dot{h}_{vu}-
(\omega_1+\mu h_1)\dot{h}_{v1}>
$$
$$
+<\omega_{v1}\omega_{1u}+h_{v1}h_{1u}-\omega_1\omega_{vu}-h_1h_{vu}>+
$$
$$
+<\omega_u{\cal{C}}^u+\omega_{uv}{\cal{C}}^v+\omega_{u1}{\cal{C}}^1+
h_u{\cal{D}}^u+h_{uv}{\cal{D}}^v+h_{u1}{\cal{D}}^1>,\eqno (44)
$$
where ${\cal{C}}^a,{\cal{D}}^b$ constitute the initial set of six differential 
constraints associated with the accompanying Lagrange-multipliers 
$\omega_{ua},h_{ub}$. The role of these constraints is to provide the values of 
$\omega_{1a},h_{1a}$ in terms of the seemingly more fundamental components 
$\omega_{va},h_{vb}$. Once they are solved, instead of the initial expression
of $S_2$ in terms  of eighteen covariant variables $\omega_{pa},h_{qb}$ we
still
have a reduced  version in terms of, at most, six independent elements: 
$\omega_{va},h_{vb}$.

It is, again, convenient to introduce $\wh{\omega}_v,\wh{\omega}_{v1},
\wh{\omega}_{1v}$
$$
\wh{\omega}_v{}''\cdot \equiv\omega_v\  \ ,\  \ \wh{\omega}_{v1}{}'\cdot \equiv\omega_{v1}
\  \ ,\  \ \wh{\omega}_{1v}{}'\cdot \equiv\omega_{1v},\eqno (45)
$$
and similar ones $\wh{h}_v,\wh{h}_{v1},\wh{h}_{1v}$ for 
the basic spin-2 field.

In terms of them, one can solve ${\cal{C}}^a=0={\cal{D}}^b$ since they
constitute  a set of the ordinary linear differential equations in the
LF-``space-like''  variable $v$. Doing so one obtains
$$
{\cal{C}}^a=0\sim \wh{h}_{1v}=\wh{\omega}_v+\partial_1\wh{h}_v\  \ ,\  \ 
h_{1u}{}'=\partial_1h_{vu}+\omega_1-\omega_{vu}\  \ ,\  \ 
h_1=\wh{\omega}_{1v}+\partial_1\wh{h}_{v1},\eqno (46a,b,c)
$$
and
$$
{\cal{D}}^b=0\sim \wh{\omega}_{1v}=(\partial_1\mu )\wh{\omega}_v+\wh{h}_v
\  \ ,\  \ \omega_{1}=\partial_1\wh{\omega}_{v1}+(-\mu\partial_1+\mu^2+1)
\wh{\omega}_v+(\partial_1-\mu )\wh{h}_v,
$$
$$ 
\omega_{1u}{}'=\partial_1\omega_{1v}+(\mu\partial_1-1){h}_{vu}+
h_1-\mu h_{1u}{}'.\eqno (47a,b,c)
$$

The covariant analysis suggest that the system has two physical relevant 
independent (spin-2$^\pm$) excitations. Six are too many.

The interesting point is that there are two additional Lagrange multipliers 
in the intermediate reduced form of $S_2(\omega_{va},h_{vb})$. Neither of
$\omega_{uv},h_{vu}$ appear in the dynamical germ  $S_2(\omega_{va}h_{vb})$,
but they are only present, linearly, in the  $LF$-generator of the dynamical
evolution, i.e. 
$$
G=<\wh{\omega}_{v1}\omega_{1u}{}'|_{\omega_{vu}=0=h_{vu}}+\wh{h}_{v1}
h_{1u}{}'|_{\omega_{vu}=0=h_{vu}}+\omega_{vu}{\cal{H}}_a+h_{vu}{\cal{H}}_h>
.\eqno (48)
$$
It turns out that the solution of these last two differential constraints 
${\cal{H}}_\omega =0={\cal{H}}_h$ is rather simple and can be cast in the form
$$
\wh{\omega}_{v1}=\wh{\omega}_{1v}\  \ ,\  \ \wh{h}_{v1}=\wh{h}_{1v}.
\eqno (49a,b)
$$

Since $\wh{\omega}_{1v}\wh{h}_{1v}$ are given in terms of 
$\wh{\omega}_{v},\wh{h}_{v}$, insertion of the information 
contained in eqs. (49a,b) leads to the final unconstrained form of $S_2$, 
expressed in terms of the two independent variables 
$\omega_{v},h_{v}$. One gets for $S_2$ (we abandon hats notation)

\vspace{3mm}
$$
S_2(\omega_v,h_v)=<-\omega_v2(1+\mu^2)\omega_v{}'{}^{\cdot}-2h_vh_v'{}^{\cdot}+
4\mu h_v\omega_v'{}^{\cdot}-
$$
$$
-\{ h_v(1+\mu^2-\partial_1^2)h_v+\omega_v[(1+\mu^2)^2+\mu^2-(1+\mu^2)
\partial_1^2]\omega_v+2h_{v}\mu[\partial_1^2-\mu^2-2]\omega_v\} >.
\eqno (50)
$$
\vspace{3mm}

\noindent
This quadratic expression can be written in a simpler way if one shifts 
${h}_v \to \widetilde{h}_v=h_v-\mu \omega_v$. $S_2$ 
becomes the almost diagonal expression (forgetting the tilde in 
$\widetilde{h}$)
$$
S_2(\omega_v,h_v)=<2\omega_v'\omega_v^\cdot +2h_v'h_v^\cdot-
\{h_v(1-\partial_1^2)h_v+\omega_v(-\partial_1^2)\omega_v+(\omega_v-\mu
h_v)^2\}>. \eqno (51)
$$

We observe that the $LF$-generator is clearly non negative. Also, as a final
check, independent variactions of $\omega_v,h_v$ yield a  system of two coupled
equations, whose inverse propagator can be quickly  computed. It turns out to be
(we have set the Fierz-Pauli mass equal to 1) $$
\Delta (\omega_v,h_v)=\square^2-(\mu^2+2)\square +1\equiv 
\Delta_+(X)\Delta_-(X)\eqno (52)
$$

In terms of the mass $X=\square^{1/2}$ we introduced before. This quartic 
polynominal has two positive 
$m^\pm =\mp 2^{-1}\mu +(1+4^{-1}\mu^2)^{1/2}$ and two additional negative 
roots $-m^\pm$.

Our original covariant action $S_2=E+T-FP$ represents a physical system of 
two decoupled spin-2 physical excitations having different masses $m^+,m^-$ 
whose $LF$-``energy'' is bounded. (Of course, the same results arise if one
performs a canonical $2+1$ analysis of $S_2$ [8]).

In terms of the two initial masses of $S_2$ (eq.39), $m^\pm$ are determined by
$$
m^\pm =\mp 2^{-1}\mu +(4^{-1}\mu^2+m^2)^{\frac{1}{2}}\eqno (53)
$$

Considering the system as the result of having spontaneously broken down the 
exact, curved TCS curved gravity, (which has one excitation of mass $\mu$) 
we observe that this spin-2 situation is exactly the uniform generalization 
of the spontaneous breakdown of the vector case [6]. The consequence of the
presence of the Fierz-Pauli mass term is the creation of the  new spin-2
excitation having the smaller mass  $-2^{-1}|\mu |+(4^{-1}\mu^2+m^2)^{1/2}$
while the previously existing  excitation of mass $|\mu |$ has been shifted to 
$2^{-1}|\mu |+(4^{-1}\mu^2+m^2)^{1/2}$.

We conclude emphasizing the much different behaviour of $LCS$ and $TCS$ 
gravities. The former cannot be spontaneously broken by the presence of 
neither a $TCS$-term nor a Fierz-Pauli one, while the already Lorentz broken, 
diffeomorphism invariant, $E+T$ action can still be broken down one 
step further by  addition of a Fierz-Pauli mass term.

\newpage

\begin{center}
{\bf REFERENCES}
\end{center}

\vspace{1cm}

\begin{enumerate}
\item S. Deser and R. Jackiw Ann. Phys. {\bf 140} (1982) 372; (E) {\bf 185} 
(1988) 406.

\item S. Deser and Z. Yang, Class. Quantum Grav. {\bf 7} (1990) 1603.

\item I. Vuorio, Phys. Lett. {\bf B163} (1985) 91; R. Percacci, P. Sodano and
I. Vuorio, Ann. Phys. {\bf 176} (1987) 344; C. Aragone, Class. Quantum Grav. {\bf 4}
(1987) L1; G. Cl\'ement, Class. Quantum Grav.{\bf 9} (1992) 35.

\item J. Horne and G. Horowitz, Nucl. Phys. {\bf B368} (1992) 444.

\item M Ba\~nados, C. Teitelboin and J. Zanelli, Phys. Rev. Lett. {\bf 69} 
(1992) 1849.

\item C. Aragone, P. J. Arias and A. Khoudeir, Il Nuovo Cimento {\bf 109B} (1994) 303.

\item C. Aragone and A. Khoudeir, Phys. Lett. {\bf B173} (1986) 141.

\item C. Aragone, P. J. Arias and A. Khoudeir, Proceedings Silarg 7, edited 
by M. Rosenbaum {\it et al}, World Scientific (Singapure) 1991 p 437.
\end{enumerate}
\end{document}